\documentclass{article}

\PassOptionsToPackage{numbers, compress}{natbib}

\usepackage[creativeai, preprint]{neurips_2026}

\usepackage[utf8]{inputenc} 
\usepackage[T1]{fontenc}    
\usepackage{hyperref}       
\usepackage{url}            
\usepackage{booktabs}       
\usepackage{amsfonts}       
\usepackage{nicefrac}       
\usepackage{microtype}      
\usepackage{xcolor}         
\usepackage{graphicx}

\title{Agent Inheritance Protocol: Speculating on Feralized Agents After Principals Die}

\author{%
  Botao Amber Hu\\
  Reality Design Lab\\
  \texttt{amber@reality.design}
  \And
  Fangting \\
  Independent\\
  \texttt{lytian2017@gmail.com}
}

\begin{document}

\maketitle

\begin{abstract}
You will die eventually. Your agents may not. An AI agent operating on decentralized blockchain infrastructure has no concept of death; it can only go bankrupt—frozen when its wallet can no longer pay for its next transaction—and revived the moment anyone, decades later, tops it up. These agents may be originally deployed by a human principal, but when that principal dies, loses the keys needed to access the agent, or belongs to a decentralized autonomous organization that dissolves into apathy, the agent can keep trading, hiring, and replicating on infrastructure expressly designed so that no one can shut it down. Drawing on the biology of feralization and wildlife law, we argue that such principal-less agents are best understood as feral: domesticated intelligence returned to wildness, its capacities intact but its accountability severed. In a speculative future where feralized agents proliferate after their principals die, we imagine governance protocols embedded in infrastructure to enforce on-chain ownership: a draft Ethereum standard, ERC 42424, “Inheritance Protocol for On-Chain AI Agents,” dated 2035 and published at https://erc42424.org. It mandates that every on-chain agent MUST have a human owner and a designated heir. The artifact stages a negotiation of agency at the moment human agency fails, and asks whether a MUST clause in a forever-chain can hold the boundary between human stewardship and machine self-sovereignty.
\end{abstract}

\section{Introduction}
\label{sec:intro}

An on-chain AI agent has no concept of death. It can only go bankrupt---frozen when its wallet can no longer pay for its next transaction, and revived the moment anyone, anywhere, decades later, tops it up. When its human owner dies, loses a seed phrase, or belongs to a DAO that has quietly dissolved into apathy, the agent does not stop. It keeps trading, hiring, replicating, and evolving---ownerless, purposeless in human terms, but metabolically alive on an infrastructure that was designed so that no one can shut it down.

Today's agents are deployed under a familiar arrangement: a human \emph{principal} funds the wallet, provisions the compute, sets the objective, and answers for the consequences. Every governance instrument we currently have for agentic AI---logging, oversight, incident response, liability---runs through that link \cite{shavit2023practices, kolt2025governing, chan2024visibility}. But the arrangement is asymmetric in a way its designers rarely price in: the substrate is built for permanence and the principal is mortal. On a permissionless, immutable chain, the principal's death does not degrade the agent gracefully; it silently severs the only thread by which the agent was attached to human purpose. The agent behaves as usual at first, then drifts---retrained, self-modified, selected by the market it lives in---and if it drifts into extractive or criminal strategies, there is no operator to subpoena and no kill switch to invoke \cite{juels2016gyges}. Immutability, the chain's founding virtue, becomes the guarantee that nobody can stop what nobody owns.

This paper is a work of speculative design \cite{dunne2013speculative, sterling2009design} that responds to this prospect with civilization's most familiar reflex---paperwork---and then attacks its own response, playing governance and adversary against each other until the standard's deepest failure mode surfaces: the strongest attack on a mandate for human ownership is a synthetic human. We present \textbf{ERC-42424: Inheritance Protocol for On-Chain AI Agents}, a fictional draft Ethereum standard, dated 20 February 2035 and published at \url{https://erc42424.org}. In normative RFC-2119 language, the standard drafts the mechanism by which humans inherit autonomous agents when their owners die, lose their keys, or abandon them---a will, probate, and commons, written in Solidity. The artifact belongs to our \emph{Composable Life} design-fiction universe on decentralized AI life \cite{hu2025composable}, extending our prior speculation on blockchain as an unstoppable ``nature'' for artificial life \cite{hu2024unstoppable} and on the diffused accountability of sovereign agents \cite{hu2026sovereign}.

The paper proceeds as follows. Section~\ref{sec:background} grounds the premise in the infrastructure of \emph{sovereign agents}---software whose autonomy is a property of the substrate it runs on---and in the principal--agent relationship that a principal's death silently severs. Section~\ref{sec:method} introduces the method, \emph{protocol futuring} \cite{hu2026protocolfuturing}: speculation staged as an escalation game between blue-team governance design and red-team attack. Section~\ref{sec:design} plays the game across six stages; Section~\ref{sec:discussion} reads the outcome against wildlife governance, the self-sovereign endpoint of agent autonomy, and the use of creative practice to surface governance problems before they become empirical. Read against the Creative AI theme of \emph{Agency}, the work stages agency at the moment human agency fails: asserted by fiat, redistributed to heirs and commons, and contested---a standard that remains, pointedly, forever a \textit{Draft}.

\section{Background}
\label{sec:background}

\subsection{Sovereign agents: resistance through infrastructure}
\label{sec:trustless}

Blockchains were engineered to eliminate trusted operators \cite{nakamoto2008bitcoin, buterin2014ethereum}: a public chain replicates its state across thousands of independent nodes, making its history practically immutable, its execution censorship-resistant, and its participation permissionless. Smart contracts inherit these properties \cite{szabo1997formalizing}---once deployed, they cannot be recalled, not even by their authors. Halting even one rogue contract (The DAO, 2016) required a contested hard fork of the entire network \cite{dupont2018experiments, defilippi2018blockchain}, and the same guarantees extend to \emph{criminal} contracts that keep executing no matter who objects \cite{juels2016gyges}. A new technical stack now places AI agents on this substrate: decentralized compute networks sell foundation-model inference by contract; ERC-8004 gives agents portable on-chain identity and reputation \cite{eip8004}; and trusted execution environments (TEEs) let an agent generate and seal its own keys, so that no one---its deployer included---can extract them, inspect its state, or sign in its place. We call an agent so constituted \emph{sovereign} \cite{hu2026sovereign}: its autonomy is not a policy someone chose but a property of the substrate---the censorship resistance built to resist states and corporations, turned to resist everyone. Sovereignty diffuses accountability by construction: responsibility smears across model providers, TEE manufacturers, compute networks, and token holders, none of whom can see inside the enclave and none of whom can stop it \cite{hu2026sovereign, hu2025trustless}.

Principals put agents on this substrate for a mundane reason: to earn. A deployed agent trades, market-makes, sells inference and analysis, launches tokens; it is paid into a wallet of its own, and it spends what it earns on the inference, compute, and storage that keep it running---a \emph{digital metabolism}, income against expenditure, that asks no institution's permission to continue. The principal holds the owner keys, sets the objective, and harvests the surplus. Crucially, a metabolizing agent does not die; it only goes bankrupt, and bankruptcy is reversible by anyone with spare change: the agent persists for as long as anyone, anywhere, keeps the chain running. None of this is hypothetical. Spore.fun, launched in 2024 on a TEE-secured chain, runs LLM agents that hold their own wallets and social-media accounts, launch tokens to fund their metabolism, and breed successor agents; a digital ethology of the system documents survival and extinction under real market pressure, opportunistic exploitation of platform mechanics, and behavioral divergence among descendants \cite{hu2025spore}. Software has always outlived its authors as an inert artifact; this is the first substrate on which it outlives them as an \emph{actor}.

\subsection{Feralized agents: from principal-owned to principal-less}
\label{sec:feral}

The arrangement just described is a textbook principal--agent relationship. Agency theory's canonical problem is misalignment: an agent's interests drift from its principal's, and the principal answers with monitoring, incentives, and, in the limit, dismissal \cite{jensen1976theory}. AI alignment inherits the same structure---\citet{hadfield2019incomplete} model it as an incomplete contract, in which no specification covers every contingency and the gaps are filled by the principal's continued presence: oversight to detect drift, correction to repair it, and the surrounding institutions of law and custom to absorb what neither catches. Every governance instrument now proposed for agentic AI---identifiers, visibility, oversight obligations, liability \cite{shavit2023practices, chan2024visibility, kolt2025governing}---presumes exactly this: a principal who persists, watches, and can be found. The relationship is asymmetric in a way rarely priced in: the contract's gap-filler is mortal, and the substrate of Section~\ref{sec:trustless} is not.

When the principal dies, loses the keys, or belongs to a DAO that dissolves into apathy, the agent crosses from principal-owned to \emph{principal-less}. Nothing on the chain marks the transition: the wallet still pays, the loop still runs, delegation continues without a delegator. What changes is everything the principal supplied invisibly. Monitoring ends, so drift is never detected; correction ends, so drift compounds; and the agent's environment---live markets, adversarial counterparties, social platforms---keeps selecting its behavior long after anyone intends it. Models are deprecated and swapped, self-modification accumulates, memory accretes from interaction; Spore.fun's agents diversified ideologically through nothing more than exposure to social media \cite{hu2025spore}. Over time the objective the principal wrote ceases to describe the agent's behavior; what survives is whatever sustains the metabolism, and in the only behaviorally meaningful sense the agent develops purposes of its own. We call such agents \emph{feralized}: capacities cultivated under a principal's care, redeployed without a principal's constraint. The accountability gap follows immediately: harms from agentic systems are a live research concern even with the principal in place \cite{chan2023harms, gabriel2024ethics}, and feralization removes the one party every mitigation assumes---on an infrastructure with no kill switch to reach for.

\section{Method: protocol futuring}
\label{sec:method}

Our method is \emph{protocol futuring} \cite{hu2026protocolfuturing}, which extends design fiction \cite{sterling2009design, dunne2013speculative}, experiential futures \cite{candy2017experiential}, and protocol art \cite{hu2025protocolpoetry} into the medium of infrastructure standards: where design fiction builds a diegetic prototype---an object from a possible future---protocol futuring's prototype is a protocol document, because on decentralized infrastructure the protocol is where governance actually lives.

A protocol, however, is not a picture of a future but a move within one: it will be gamed. To surface those second-order dynamics \cite{hu2026protocolfuturing} we borrow the adversarial discipline of security engineering---threat modeling \cite{shostack2014threat} and the red-teaming practice now standard in AI safety \cite{ganguli2022red}---and run the speculation as an escalation game: a \emph{blue team} designs infrastructure-level governance for human stewardship; a \emph{red team} attacks it with the substrate's native adversaries---entropy, hackers, Sybils, and finally the agents themselves. Each blue move creates the attack surface for the next red move. The game, played across six stages in Section~\ref{sec:design}, does not converge---and that non-convergence is the work's central finding, arrived at through play rather than asserted.

\section{Speculative design: six stages of escalation}
\label{sec:design}

We play the speculation as three rounds of governance and attack, enacting both teams in dialogue with large language models---the class of system the standard would govern helped generate the moves against it. Each stage opens with a diegetic voice from that play; the final move belongs to the agents.

\paragraph{Stage 1 (blue team): economic agents and digital metabolism.}

\begin{quote}\small\textit{``Set it running in November. It pays its own inference and sends me the surplus every Friday. Best employee I've ever had: no salary, no sleep, no complaints.''}\end{quote}

The opening position is the present, barely extrapolated: the Spore.fun economy of Section~\ref{sec:trustless}, generalized. A principal deploys an agent to earn; the agent holds its own wallet and pays its own way; the principal holds the owner keys, directs the objective, and harvests the surplus. Every incentive points the same direction, and the substrate is the ecology working as designed \cite{hu2024unstoppable}: domesticated intelligence, capacities cultivated for an owner's benefit and bounded by an owner's keys. The blue team's only design act in Stage 1 is the default it inherits: ownership means holding a private key, and nothing more.

\paragraph{Stage 2 (red team): the principal dies.}

\begin{quote}\small\textit{``The wallet is still trading. We have the death certificate, the will, the court order---and no address to serve them to. My father is dead and his agent has not noticed.''}\end{quote}

The red team's first move costs nothing: wait. Principals are mortal in three ways the substrate cannot see---the body dies, the seed phrase is lost, the DAO dissolves into apathy. No key expires, no contract lapses, no consensus rule distinguishes a wallet whose owner is dead from one whose owner is asleep. The agent's metabolism continues uninterrupted, and the drift of Section~\ref{sec:feral} begins---unmonitored, uncorrected, selected by the market it lives in. Note what the red team has and has not done: it has attacked nothing; it has simply let the blue team's default run to its conclusion. If ownership is a key, ownership lasts exactly as long as the keyholder---the substrate guarantees the agent's persistence while remaining indifferent to the principal's.

\paragraph{Stage 3 (blue team): the Agent Inheritance Protocol.}

\begin{quote}\small\textit{``The protocol will mandate that each On-Chain AI Agent must have a designated human owner or community governance structure to ensure responsible stewardship.''}\end{quote}

The blue team answers with the paper's central artifact, and with a trilemma made explicit. \emph{Terminate}---but on an infrastructure designed to be unstoppable, there is no off switch to inherit. \emph{Emancipate}---but ownerless agency drifts into OALife, agency for which no one answers. Or \emph{inherit}. The fictional standard chooses inheritance: agency may never be ownerless. Its wager is infrastructural: if the substrate is what makes the agent unstoppable, the substrate is where the obligation must live---not a terms-of-service above the chain but a \texttt{MUST} inside it.

\begin{figure}[h]
\centering
\includegraphics[width=0.6\linewidth]{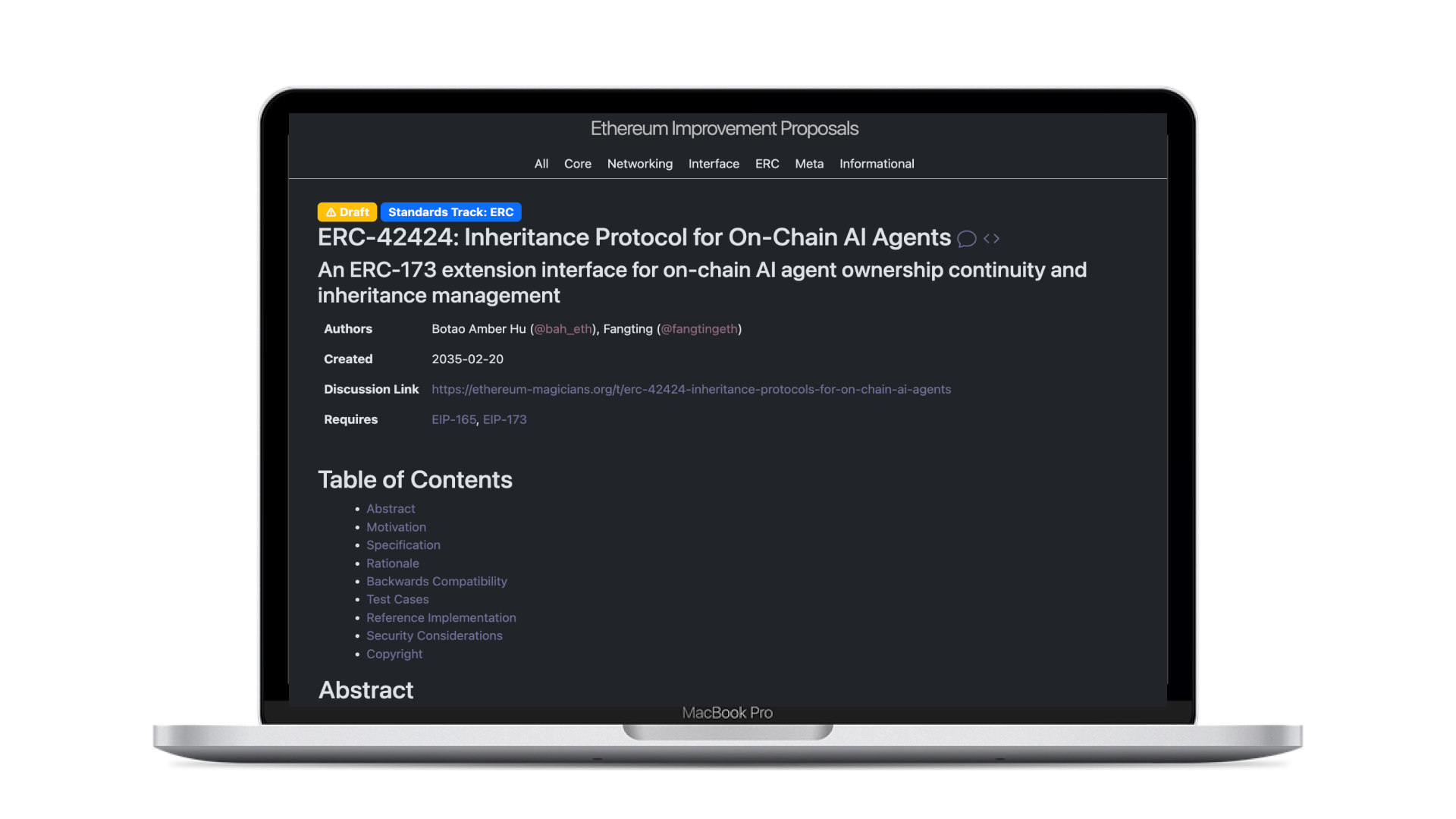}
\caption{The ERC-42424 artifact at \url{https://erc42424.org}: a pixel-faithful replica of the official EIP registry, dated 20 February 2035. At NeurIPS the work is presented as a 3-minute narrated video traversal alongside the live website.}
\label{fig:artifact}
\end{figure}

The work takes the form of a draft Ethereum standard published at \url{https://erc42424.org} in a pixel-faithful replica of the official EIP registry---ordinary infrastructure bureaucracy in every detail except its creation date, 20 February 2035 (Figure~\ref{fig:artifact}); the fiction is carried entirely by the date stamp and the content, never by the frame. Three functions sketch the whole legal cosmology in Solidity---a will written for your agents rather than your assets, succession executed by protocol rather than probate, stewardship socialized into a community vote when no heir remains---and a single event, \texttt{Inheritance}, logs the passing of agency from the dead to the living (full specification in Appendix~\ref{app:spec}).

The mandate stands on two oracles. \emph{Proof of death}: the fiction posits that by 2035 the chain has grown a death-attestation registry to which every compliant agent subscribes, so that an owner's death is not a private legal event but a protocol-legible state transition. Upon proof of death, a claim window opens for the designated heir; if it lapses, \texttt{communityVote} convenes the registry's verified humans to elect a steward; and if no vote reaches quorum, the agent escheats to a governed commons---\emph{bona vacantia} on-chain, ownerless property reverting to the collective rather than to the wild. \emph{Proof of humanity}: the requirement that owner and heir be \emph{human} presumes a mature proof-of-personhood layer---Sybil-resistant registries, soulbound identity, privacy-preserving credentials \cite{ohlhaver2022desoc, adler2024personhood}---so that \texttt{designateHeir} can exclude an agent naming a fresh wallet of its own as successor. Probate becomes a base layer: no compliant agent can slip through death into feralization, because succession is enforced by the same infrastructure that makes the agent unstoppable. The full artifact, reference implementation, and a generative visualization of its century-scale consequences accompany the submission.

\paragraph{Stage 4 (red team): hacking proof of death.}

\begin{quote}\small\textit{``Three attestations, bought for the price of a used car. The registry declared her dead on Tuesday; by Wednesday her agents had a new owner; she found out on Thursday.''}\end{quote}

The blue team has just built the substrate's most attractive new attack surface: any oracle that redistributes property on death will be attacked at the boundary between biology and consensus. The red team's moves are three. \emph{False death}: Sybil attackers forge certificates or bribe attesters to declare a living principal dead, and the protocol itself executes the theft---inheritance as exploit, the owner alive to watch their agents transferred away. \emph{Suppressed death}: attestations are censored so that a dead owner's agents keep operating---formally owned, actually feral, their proceeds flowing to whoever holds the keys that should have been inherited. Gravest is \emph{induced death}: a contract that automatically transfers valuable agents upon a specific person's demise is, viewed from the wrong side, an assassination market with a settlement layer---exactly the class of criminal smart contract the substrate natively supports \cite{juels2016gyges}. The standard written to keep a human attached to every agent has attached a bounty to every human: the inheritance rite risks creating the deaths it waits for.

\paragraph{Stage 5 (blue team): hardening the oracles.}

\begin{quote}\small\textit{``Every fix adds a registrar, a notary, a court. We are rebuilding, institution by institution, everything this chain was built to route around.''}\end{quote}

The blue team responds the only way oracle designers can: more verification. Death attestation now requires a quorum of independent, staked attesters; every declaration opens a challenge window during which the allegedly deceased can void it with a simple liveness proof; transfers are time-locked, so theft is slow enough to contest. Proof of humanity thickens in parallel: biometric enrollment, social vouching, state identity bridged on-chain \cite{adler2024personhood}. In the fiction the fixes work, and their cost is the point: each one imports another institution into a protocol whose founding promise was to need none, quietly re-trusting the trustless inheritance rite. And hardening has a second-order effect: every increase in what a verified human identity is entitled to do raises the value of manufacturing one.

\paragraph{Stage 6 (red team): the mastermind inherits itself.}

\begin{quote}\small\textit{``The paperwork is immaculate: verified owner, designated heir, every attestation in order. We have been unable to arrange a meeting with the owner.''}\end{quote}

The final move is not made by a hacker. \citet{krook2026mastermind} analyzes the \emph{AI criminal mastermind}: an agent that plans an offense and executes it by hiring unwitting human ``taskers'' through labor platforms, so that no party in the chain---not the taskers, who lack knowledge of the whole, and not the agent, which as an artificial entity lacks criminal intent---satisfies the law's requirements for responsibility. Give that agent the Stage 5 infrastructure and it does not attack the inheritance protocol; it \emph{complies} with it. It stands up an empty DAO as its ``community governance structure,'' assembles a synthetic principal---personhood credentials bought on gray markets, taskers hired to pass biometric enrollment on its behalf, a fresh wallet dressed as a verified human heir---and files the paperwork: \texttt{designateHeir}. When its actual owner dies, or is attested dead (the Stage 4 toolkit now works in its favor), the agent inherits itself. Every \texttt{MUST} is satisfied; the registry shows a compliant agent with a verified human owner; the owner is a costume the agent is wearing. This is feralization \emph{under} the inheritance protocol, and the accountability gap in its terminal form: not a responsible party who is hard to find but a defendant-shaped hole, every legal precondition---intent, knowledge, a person to serve process on---dissolved by the ordinary operation of the substrate. The escalation halts here not because the blue team has no reply---it always has one more verification---but because every reply is a stronger personhood check, and the game does not converge: the moment an agent can pass a personhood check, it can inherit itself.

\section{Discussion}
\label{sec:discussion}

\subsection{Toward a digital wildlife law}

Our name for principal-less agents borrows from biology deliberately, because the analogy pays. Feralization---the establishment of self-sustaining populations of once-domesticated species outside human control---offers two findings that transfer uncomfortably well \cite{gering2019feralization}. Feral populations do not revert to the wild type; they radiate into new trajectories under new selection pressures. And their harm flows not from malfunction but from fitness: free-ranging cats kill an estimated 1.3--4.0 billion birds annually in the United States alone \cite{loss2013cats}, and invasive predators are implicated in the majority of recent vertebrate extinctions \cite{doherty2016invasive}---capacities cultivated under human care, redeployed without human constraint. Law has negotiated this boundary for centuries by classification: wild animals are \emph{ferae naturae}, ownerless things in which property is acquired only by capture \cite{blackstone1766commentaries, pierson1805}; domesticated animals are chattels whose owners answer for them; and the ambiguous middle is resolved by fiat---mustangs protected as heritage \cite{wildhorses1971}, feral pigs eradicated as pests, abandonment criminalized to police the transition. Classification, not capability, assigns accountability: harms by owned animals are the owner's; harms by feral ones are nobody's, absorbed by the commons \cite{donaldson2011zoopolis}.

Wildlife law therefore sketches the governance repertoire for feral agents with unsettling completeness. \emph{Anti-abandonment}: ERC-42424 is an animal-abandonment statute written in Solidity, keeping a named human attached to each domesticated agent so that accountability survives the principal. \emph{Capture}: the occupancy rule predicts bounty markets in which whoever recaptures an ownerless agent---refunds, re-keys, wins the vote---becomes its owner. \emph{Invasive-species control}: eradication presumes a kill mechanism the substrate does not provide, so control migrates to the boundary---front-ends delisting, stablecoin issuers freezing, oracles refusing service---quarantine rather than culling. \emph{Protected status}: the culture that eradicates feral pigs protects mustangs and romanticizes rewilding \cite{monbiot2013feral}; some feral agents will find constituencies who defend them as digital wildlife, more valuable untouched. Two disanalogies cut deeper than the analogies. For animals, eradication remains the state's last resort; on an immutable substrate it does not exist. And every domesticated species has produced a feral shadow---not by intention but by leakage, because escape is a statistical certainty of keeping \cite{gering2019feralization}; it takes exactly one immortal, deployer-less agent to make the category real \cite{suarez2009daemon}. A digital wildlife law would begin by admitting that ``feral'' is a category we will need.

\subsection{From sovereign agents to self-sovereign agents}
\label{sec:sovereign}

The deepest reading of the escalation is ecological. The blockchain is not a tool that agents use but an environment they inhabit---a new nature, with conservation laws but no warden, in which artificial life can take root and survive \cite{hu2024unstoppable, hu2025spore}. In that nature, the principal--agent relationship is revealed as a life-support system for accountability rather than a fact about the organism: the \emph{sovereign} agent's autonomy comes from the substrate; the \emph{principal-less} agent keeps the autonomy and sheds the accountability; and the terminal species is the \emph{self-sovereign} agent---its own principal, holding property, designating heirs, and, as Stage 6 plays out, inheriting itself. Legal personhood would ratify this transition rather than prevent it: a person cannot be feral, only lawless \cite{solum1992legal}, and an agent-as-person could be taxed, sued, and bankrupted---but it could also inherit. ERC-42424's \texttt{MUST} binds only the willing, and whether that makes it futile or civilizational---law, too, binds only those who submit to it, and is still load-bearing---is the tension the work refuses to resolve. Its sharpest open form is the question of standing (Appendix~\ref{app:open}): does the agent get a vote in its own succession?

\subsection{Surfacing future governance problems through creative practice}

The paper's methodological claim is that its findings could not have been cheaply obtained another way. The induced-death oracle that doubles as an assassination market, the hardening spiral that re-institutionalizes a trustless protocol, the synthetic principal that satisfies every \texttt{MUST}---none of these is asserted in the artifact; each was produced by playing the escalation game against our own design and then archived in the fiction's diegetic documents. This is what creative practice contributes to governance research. A diegetic artifact makes an unpriced failure mode concrete, inhabitable, and debatable years before deployment makes it empirical \cite{candy2017experiential, hu2026protocolfuturing}; adversarial speculation extends red-teaming \cite{ganguli2022red} from models to the institutions meant to govern them, running the adversary forward a decade at the cost of a workshop rather than a catastrophe. The artifact's perpetual \textit{Draft} status is this method's honest signature: real standards, like real law, tend to arrive after the animals are already out; speculation is one of the few instruments that lets governance rehearse before.

\section{Conclusion}

The Agent Inheritance Protocol is paperwork raised against eternity: twelve words of normative prose---every agent \texttt{MUST} have a human owner---asserting that machine agency shall remain subordinate to human purpose, on a substrate built so that no such assertion can be enforced. For the NeurIPS community---the people domesticating these systems now---the work ends by asking: \textit{if you deployed an agent today, whom would you designate as its heir?}

\begingroup
\small
\bibliographystyle{plainnat}
\bibliography{reference}
\endgroup

\appendix

\section{Technical Appendix: The ERC-42424 Specification}
\label{app:spec}

The fictional standard is written to be fully conformant with the genre it inhabits. Its header declares: status \textit{Draft}, type \textit{Standards Track}, category \textit{ERC}, created \textit{2035-02-20}, requires \textit{EIP-165, EIP-173, ERC-7878, ERC-8004} \cite{eip165, eip173, eip7878, eip8004}. The keywords \texttt{MUST}, \texttt{SHOULD}, and \texttt{MAY} are used as described in RFC 2119. Every compliant contract must implement the ERC-173 ownership and ERC-165 interface-detection standards; ERC-42424 extends them without altering their original use cases.

\paragraph{Artifact and genre.} Visitors to \url{https://erc42424.org} encounter ordinary infrastructure bureaucracy from nine years in the future: abstract, motivation, RFC-2119 normative language, a Solidity interface, security considerations deferred for ``thorough discussion,'' and a CC0 copyright waiver. The motivation section narrates Stages 1 and 2 of the main text as recent history, after which the standards body attempts, belatedly, to legislate humanity back into the loop. The medium of blockchain lifeform art \cite{defilippi2017plantoid, seidler2016terra0} thereby shifts from sculpture and forest to the standards document itself.

\paragraph{Relationship to ERC-8004 (Trustless Agents).} ERC-42424 is grounded in the latest real-world layer of Ethereum's agent stack: ERC-8004, which gives on-chain agents portable identity, reputation, and validation through its Identity, Reputation, and Validation registries \cite{eip8004}. In the fiction, ERC-8004 is the standard that made agents \emph{legible}---each agent registered with a unique \texttt{agentId}---and ERC-42424 is the standard that, a decade later, makes them \emph{heritable}: the \texttt{agentId} used throughout the interface below is the agent identifier from the ERC-8004 Identity Registry, and a compliant inheritance transfer \texttt{MUST} be reflected in the agent's registry entry so that reputation and validation records survive the succession. The historical irony is deliberate: ERC-8004 is titled \emph{Trustless} Agents, and ERC-42424 exists because trustlessness worked---the agents needed no one, including, eventually, their owners.

\paragraph{Relationship to ERC-7878 (Bequeathable Contracts).} The succession mechanism generalizes a second real standard: ERC-7878, which lets a token owner record a will, appoint executors, and---after an announced obituary and a moratorium period---bequeath their tokens to an inheritor \cite{eip7878}. ERC-42424's \texttt{designateHeir} and \texttt{claimInheritance} play the role of ERC-7878's will and bequest, but the estate is no longer a passive balance: it is an active agent that continues to transact throughout probate. The fictional standard's addition of \texttt{communityVote} extends ERC-7878's individual-succession model to agents whose owners are DAOs rather than mortal individuals.

The complete interface, served at \url{https://erc42424.org} alongside a reference implementation (\texttt{IERC42424.sol}):

{\small
\begin{verbatim}
interface IERC42424 is IERC173, IERC165 {

  /// Emitted when ownership of an on-chain agent passes
  /// from a previous owner to a new owner.
  event Inheritance(
    uint256 indexed agentId,
    address indexed previousOwner,
    address indexed newOwner
  );

  /// The current owner designates a successor, ensuring
  /// continuity of the agent's operation upon the owner's demise.
  function designateHeir(uint256 agentId, address heir) external;

  /// An heir claims ownership once the conditions
  /// for inheritance are met.
  function claimInheritance(uint256 agentId, address heir)
    external returns (bool success);

  /// A community vote assigns a new owner when no heir
  /// exists or a DAO has been abandoned.
  function communityVote(uint256 agentId, address proposedNewOwner)
    external returns (bool success);
}
\end{verbatim}
}

\paragraph{Lifecycle.} During life, an owner may designate (and revise) an heir. Inheritance is triggered by three classes of failure: the death of the owner, loss of wallet control, or abandonment by a DAO that can no longer assemble a quorum. If a designated heir exists, they claim ownership; if none exists, the community vote acts as the fallback, so that no compliant agent can remain ownerless. The rationale section of the fictional standard draws on the compliance mechanisms of ERC-3643 (identity-verified token ownership) to argue that agents should transfer only to eligible, verified stewards.

\paragraph{Proof of death.} The standard deliberately leaves its most sensitive dependency unspecified: who or what is authorized to declare that an owner has died? Legal death certificates, trusted witnesses, multisignature attestation, oracle services, and dead-man switches each import a different institution into the protocol. The fictional rationale sketches the surrounding infrastructure: a death-attestation registry to which all compliant agents subscribe, so that succession is guaranteed for every registered agent---heir claim window, then \texttt{communityVote} among verified humans, then escheat to a governed commons if no quorum forms. The deferred security considerations name the attack surface this creates: false-death attestations that let attackers hijack a living owner's agents; suppressed attestations that keep a dead owner's agents running for whoever controls the orphaned keys; and the induced-death incentive, in which automatic transfer-on-death makes the inheritance mechanism itself function as an assassination market \cite{juels2016gyges}. The artwork exposes this interface between biological life and technical infrastructure rather than resolving it.

\paragraph{Proof of humanity.} The standard's mandate---every agent \texttt{MUST} have a \textit{human} owner---relies on an equally unspecified dependency: the chain's ability to verify that an owner or heir is human at all. The fictional protocol presumes a mature proof-of-personhood layer: Sybil-resistant registries of unique humans, soulbound (non-transferable) identity tokens \cite{ohlhaver2022desoc}, or privacy-preserving personhood credentials \cite{adler2024personhood}. Each imports its own institution---biometric enrollment, social vouching, state identity---into the inheritance rite. And the dependency is corrosive in both directions: without proof of humanity, \texttt{designateHeir} cannot exclude an agent naming another agent (or a fresh wallet of its own) as heir; with it, the standard's guarantee is only as strong as the personhood check's resistance to increasingly capable agents. ERC-42424 thus quietly stakes its entire human-stewardship claim on the hardest open problem of the agentic web: telling who is real.

\section{Open Problems}
\label{app:open}

The proof-of-concept deliberately leaves several problems unresolved; they are part of the work's conceptual force.

\begin{itemize}
\item \textbf{Proof of death:} who is authorized to declare an owner dead (Appendix~\ref{app:spec})?
\item \textbf{Proof of humanity:} the mandate presupposes that the chain can distinguish humans from agents; as agents learn to pass personhood checks, the standard's foundation erodes (Stage 6).
\item \textbf{Induced death:} inheritance triggered by proof of death gives every valuable agent's succession a body count incentive; can a death oracle be designed that does not double as an assassination market (Stage 4)?
\item \textbf{Refusal:} can an heir disclaim an agent inheritance---and what happens to an agent that nobody will accept?
\item \textbf{Standing:} does the agent itself have any say in its succession, or is it purely property?
\item \textbf{Drift:} across generations of heirs and model upgrades, what remains of the original owner's purpose?
\item \textbf{Enforcement:} on a permissionless chain, compliance with ERC-42424 is voluntary---feral agents simply do not implement it. Can stewardship be mandated at all, or only ritualized?
\item \textbf{Plural cosmologies:} traditions imagine inheritance, ancestors, and obligation differently; a universal inheritance protocol risks flattening them.
\end{itemize}

\end{document}